\documentstyle[12pt]{article}
\begin{document}
\input FEYNMAN
\def\be{\begin{eqnarray}}
\def\en{\end{eqnarray}}
\def\non{\nonumber}
\def\la{\langle}
\def\ra{\rangle}
\def\ep{\varepsilon}
\def\qs{q\!\!\!/}
\def\tr{{\rm tr}}
\def\ov{\overline}
\def\pr{{\sl Phys. Rev.}~}
\def\prl{{\sl Phys. Rev. Lett.}~}
\def\pl{{\sl Phys. Lett.}~}
\def\np{{\sl Nucl. Phys.}~}
\def\zp{{\sl Z. Phys.}~}

\font\el=cmbx10 scaled \magstep2
{\obeylines
\hfill CLNS 96/1393
\hfill IP-ASTP-01-96
\hfill January, 1996}
\vskip 1.5 cm

\centerline{\large \bf CKM Favored Semileptonic Decays of Heavy Hadrons}
\centerline{\large \bf at Zero Recoil}
\medskip
\bigskip
\medskip
\centerline{\bf Tung-Mow Yan}
\medskip
\centerline{Floyd R. Newman Laboratory of Nuclear Studies, Cornell University,}
\centerline{Ithaca, NY 14853, USA}
\bigskip
\centerline{\bf Hai-Yang Cheng and Hoi-Lai Yu}
\medskip
\centerline{ Institute of Physics, Academia Sinica,}
\centerline{ Taipei, Taiwan 115, Republic of China}
\bigskip
\bigskip
\centerline{\bf Abstract}
\bigskip

{\small

We study the properties of  Cabibbo-Kobayashi-Maskawa (CKM)
favored semileptonic decays of mesons and 
baryons containing a heavy quark at the point of no recoil. We first use a
diagrammatic analysis to rederive the result observed by earlier authors that
at this kinematic point the $B$ meson decays via $b\to c$ transitions can only
produce a $D$ or $D^*$ meson. The result is generalized to include photon 
emissions which violate heavy quark flavor symmetry. We show that photons 
emitted by the heavy quarks and the charged lepton are the only light 
particles that can 
decorate the decays $\bar{B}\to D(D^*) + \ell\nu$ at zero recoil,
and the similar processes 
of heavy baryons. Implications for the determinations of the CKM parameter 
$V_{cb}$ are discussed. Also studied
in this paper is the connection between our diagrammatic analysis of
suppression of particle emission and the formal observation based on weak 
currents at zero recoil being generators of heavy quark symmetry. We show
that the two approaches can be unified by considering the Isgur-Wise function
in the presence of an external source.

}

\pagebreak

\noindent{\bf I.~~Introduction}

   In semileptonic decays of heavy mesons and baryons containing a $b$ 
quark, a large phase space is available for emission of many light particles 
such as pions and photons. Synthesis of spin and flavor symmetry of heavy 
quarks and chiral symmetry of light quarks provides a natural framework to 
describe these processes involving soft pions and photons [1,2,3]. 
At zero recoil, the
Cabibbo-Kobayashi-Maskawa (CKM) favored semileptonic decays of heavy mesons and
baryons exhibit some remarkable properties. For instance, it has been 
pointed out [4,5] that within strong interactions at the point of 
no recoil, the CKM
favored semileptonic decays of a $B$ meson in the heavy quark 
limit can only produce a $D$ or $D^*$ meson. This follows from the 
observation
that when the initial and final heavy quarks have the same velocity, the weak
currents become generators of heavy quark symmetry and their action on
a $B$ meson can only give rise to a linear combination of $D$ and $D^*$
mesons. At this kinematic point, the amplitudes for CKM favored semileptonic
decays of a $B$ meson accompanied by emission of light hadrons are of order
$1/m_Q$ and the rates are suppressed by $1/m^2_Q$.

   The importance of the above result is beyond the
obvious theoretical interest of being a precise statement. It implies
that the Isgur-Wise function measured experimentally at the point of no recoil
is not contaminated by corrections of order 1 or order $1/m_Q$ due to
emissions of pions. When combined with the Luke's theorem [6], the above 
result
would imply that the corrections to the Isgur-Wise function at the point of no 
recoil are at least of order $1/m_Q^2$. Motivated 
by these observations, we devote this paper to a detailed stuty of the 
properties of CKM favored semileptonic decays of heavy hadrons at zero
recoil, including a brief discussion in the last section of measuring the
CKM parameter $V_{cb}$ in $\bar{B}\to D^*+\ell\nu$.
Although the formal argument of 
refs.[4,5] is as simple as it is, it is desirable and instructive
to have an explicit derivation of this result based on an analysis of
the multiparticle amplitudes. The reason is obvious. The amplitudes  for
$\bar{B}\to D(D^*)+\pi+\ell\nu$ vanish at $v=v'$ due to cancellation between
emissions of the pion before and after the weak vertex and cooperation
of pion emission from different intermediate states. As the number of pions 
increases, the cancellation becomes increasingly complex. One of the
purposes of the present paper is to show how to organize the diagrams
systematically to establish the general results to all orders in 
perturbation theory, not only for emission of pions and other Goldstone 
bosons, but also for emission of any other light hadrons. Another
purpose of our study is to examine if the result still holds 
in the presence of electromagnetic interactions. First of all, photon
emission from the heavy quarks violates the heavy flavor symmetry since the
$b$ and $c$ quarks have different electric charge. Secondly, the strong 
interaction coupling constant $g$ responsible for pion emission
is rather small, $g\sim 0.3-0.5$. Photon emission could, in principle, compete
favorably with pion emission. We will show that photon emission is
forbidden if it is accompanied by any other light particles. Only photons
emitted by the heavy quarks and the charged lepton need to be considered
and they are to accompany the semileptonic decays $\bar{B}\to D(D^*)\ell\nu$.
Moreover, these photon emissions can be calculated by the standard QED
technique. However, we must hasten to add that
photons treated here are not the ``soft photons'' in the conventional sense;
their momenta can be comparable with those of pions. Of course, in order for
us to make use of the framework of heavy quark effective theory, the momenta
of all light particles must be small compared with the heavy quark masses. In
principle, we can control such a kinematic region by selecting leptons with
appropriate momenta. 

   Finally, we offer a proof of the general result which relates the formal 
argument to our diagrammatic analysis. Specifically, we consider the
Isgur-Wise function in the presence of an external source which couples only 
to the light quark's degrees of freedom. The Isgur-Wise function has still the
normalization unity at the point of no recoil. This is the statement that
the weak current at this kinematic point can only change a $B$ meson into a
$D$ or $D^*$ meson. But now, all of its functional derivatives evaluated at 
zero source must vanish. These functional derivatives reproduce
the pion emission amplitudes at zero recoil.

   We should point out that the diagrammatic analysis reveals a feature
which is not apparent from the formal argument. Consider the amplitude for 
emission of $n$ pions in addition to the $D$ or $D^*$ meson. There are $n!$
sets of diagrams which differ only in permutations of the pions. Our analysis 
shows that each of the $n!$ sets vanishes by itself. An interesting and 
obvious question is why not all the $n!$ sets are needed to make the amplitudes
vanish at $v=v'$. Is there any deep physics hidden here ? At the moment, we do 
not have an answer.

   Once we have seen how the suppression
of particle emissions works at the point of no recoil for the CKM
favored semileptonic decays of heavy mesons, it is straightfoward to extend
the same analysis to the corresponding case of heavy baryons.

\vskip 0.6mm
\noindent{\bf II.~~Suppression of a Single Soft Pion Emission at Zero Recoil}

    To gain some insight from simple examples and to prepare for a 
general treatment in the next section,   
   we shall recall in this section the explicit calculations in [1,7] for
the CKM favored semileptonic weak decays of heavy hadrons involving one
soft pion emission: $\bar{B}\to D(D^*)+\pi+\ell\nu,~\Sigma_b\to\Sigma_c(
\Sigma_c^*)+\pi+\ell\nu$ and $\Sigma_b\to\Lambda_c+\pi+\ell\nu$. The coupling
strength for a single pion emission is proportional to $q/f_\pi$ where $q$ 
is the pion momentum, so the semileptonic decay amplitude for producing
a pion is naively expected not to be suppressed when $q\sim f_\pi$. However,
we shall see that the amplitudes for pion emission from the initial
hadron and final hadron conspire to give a zero sum at zero recoil in the
heavy quark limit. 

    Figs.~1 and 2 show the Feynamn diagrams for $\bar{B}\to D(D^*)\pi\ell\nu$.
The amplitudes have been evaluated in [1,7] 
with $v'\neq v$ (see Eqs.~(4.27-4.29) of [1] and Eqs.(3.3-3.8) as well
as Eqs.(4.1-4.4) of [7]). At zero recoil $v\cdot v'=1$, the amplitudes are 
given by
\be
\la D(v)\pi^a(q)|J^W_\mu|\bar{B}(v)\ra &=& F[(q\cdot v)v_\mu-q_\mu]\left({1
\over v\cdot q+\Delta_B}-{1\over v\cdot q-\Delta_D}\right),   \\
\la D^*(v,\ep')\pi^a(q)|J^W_\mu|\bar{B}(v)\ra  &=& F\bigg[{1\over -v\cdot 
q-\Delta_B}[i\epsilon_{\mu\nu\lambda\kappa}\ep'^\nu q^\lambda v^\kappa-
(\ep'\cdot q) v_\mu]   \non \\
&& +{1\over v\cdot q}i\epsilon_{\mu\nu\kappa\lambda}\ep'^\nu q^\lambda v^\kappa
-{1\over v\cdot q+\Delta_D}(\ep'\cdot q)v_\mu\bigg].
\en
In above equations, $\Delta_B=M_{B^*}-M_B$, $\Delta_D=M_{D^*}-M_D$, and
\be
F=\,iu(\bar{B})^\dagger{1\over 2}\tau_a u(D^*)\sqrt{M_BM_{D^*}}\,
{f\over f_\pi}C_{cb}\xi(1),
\en
where $u(P)$ is the isospin wave function of the heavy meson $P$,
$C_{cb}$ is a QCD correction factor, and $\xi(1)$ is the Isgur-Wise 
function evaluated at $v\cdot v'=1$.
In the heavy quark limit, $\Delta_B =\Delta_D={\cal O}(1/m_b)$ or ${\cal O}
(1/m_c)$. Therefore, we see that 
the $\bar{B}\to D(D^*)\pi\ell\nu$ amplitudes vanish at zero recoil.

   There are three baryonic Isgur-Wise form factors for weak transitions of 
heavy baryons: $\zeta(v\cdot v')$ for antitriplet-antitriple transition, e.g.,
$\Lambda_b\to\Lambda_c$, and $\xi_1(v\cdot v')$, $\xi_2(v\cdot v')$ 
for sextet-sextet transition with the normalization $\zeta(1)=\xi_1(1)=1$.
The form factor $\xi_2$ drops out at zero recoil. The 
amplitude deduced from the Feynman diagrams in Fig.~3 at $v\cdot v'=1$ reads 
(see Eqs.~(4.42) and (4.43) of [1])
\be
\la\Sigma^f_c (v, s^\prime)\pi^d (q)|J^W_\mu|\Sigma^e_b(v, s)\ra
=\varepsilon_{def}{g_1\over 2 f_\pi}C_{cb}\xi_1(1)
\overline{u} (v,s^\prime)[C_{a\mu} + C_{b\mu} + C_{c\mu} + C_{d\mu}]u(v, s),
\en
where $\varepsilon_{def}$ is the totally antisymmetric symbol
associated with the isospin of the particles involved, $g_1$
is one of the unknown coupling constants given in Eq.(40) below,
and $C_{a\mu} \cdots C_{d\mu}$
correspond to the contributions from Fig.~3(a) $\cdots$ 3(d) respectively:
\be
C_{a\mu} &=& \frac{1}{3} \cdot \frac{1}{- v \cdot q} [\gamma_\mu
(1-\gamma_5)+ 4 v_\mu \gamma_5 ] (\qs - q \cdot v ), \\
C_{b\mu} &=& \frac{1}{ v \cdot q + \Delta_{\Sigma_b}}
(1-\gamma_5) [-q_\mu + \frac{1}{3} \gamma_\mu (\qs - q \cdot v) 
+ \frac{1}{3} v_\mu (\qs + 2q \cdot v) ], \\
C_{c\mu} &=& \frac{1}{3} \cdot \frac{1}{v \cdot q} (\qs - q
\cdot v)[\gamma_\mu (1-\gamma_5) -4 v_\mu \gamma_5],  \\
C_{d\mu} &=& - \frac{1}{v\cdot q -\Delta_{\Sigma_c}}
[-q_\mu+\frac{1}{3}(\qs -q\cdot v) \gamma_\mu + \frac{1}{3}
(\qs + 2q \cdot v) v_\mu ] (1+ \gamma_5),
\en
where $\Delta_{\Sigma_{b(c)}}=M_{\Sigma^*_{b(c)}}-M_{\Sigma_{b(c)}}$.
Using the fact that $\bar{u}(v)\gamma_5u(v)=0$ and $\Delta_{\Sigma_{b(c)}}=
{\cal O}(1/m_{b(c)})$ in the heavy quark limit, we find that
$C_{a\mu}+C_{c\mu}=(\qs\gamma_\mu-\gamma_\mu\qs)/(3v\cdot q)$ is exactly
canceled out by $C_{b\mu}+C_{d\mu}$. 

   The next example is the semileptonic decay 
$\Sigma_b\to\Sigma_c^*\pi\ell\nu$ with the amplitude, for $v\cdot v'=1$ (see
Eqs.~(4.44) and (4.45) of [1])
\be
\la\Sigma^{\ast f}_c (v, s^\prime) \pi^d(q)|J^W_\mu|
\Sigma^e_b (v, s )\ra
= \varepsilon_{def} \frac{g_1}{2 f_\pi} C_{cb}\xi_1(1)\overline{u}_\lambda
(v, s') [D^\lambda_{a\mu} + D^\lambda_{b\mu}
+ D^\lambda_{c\mu} +D^\lambda_{d\mu} ] u (v, s),
\en 
where $u_\lambda(v,s)$ is a Rarita-Schwinger vector spinor, and
$D^\lambda_{a\mu} \cdots D^\lambda_{d\mu}$ are the
contributions from each of the Feynman diagrams in Fig.~4 respectively,
\be
D^\lambda_{a\mu} &=& \frac{2}{\sqrt{3}} \cdot \frac{1}{- v \cdot q}
g^\lambda _\mu (1+ \gamma_5) (\qs- q \cdot v),  \\
D^\lambda_{b\mu} &=& \frac{\sqrt{3}}{2} \cdot
\frac{1}{- v\cdot q - \Delta_{\Sigma_b}}
[- \gamma_\mu (1- \gamma_5) q^\lambda + \frac{2}{3}
g^\lambda _\mu (1+ \gamma_5)(\qs- q \cdot v) ],  \\
D^\lambda_{c\mu} &=& \frac{1}{2 \sqrt{3}} \cdot
\frac{1}{v\cdot q +\Delta_{\Sigma_c}}q^\lambda [ \gamma_\mu (1- \gamma_5) - 4
v_\mu ],  \\
D^\lambda_{d\mu} &=&\sqrt{3} \frac{1}{v\cdot q} [ - g^\lambda
_\mu ( \qs- q \cdot v) + \frac{2}{3} q^\lambda
(\gamma_\mu - v_\mu) ] (1+ \gamma_5).
\en
Noting that $\bar{u}_\lambda(v)\gamma_5 u(v)=0,~\bar{u}_\lambda(v)(\gamma_\mu
-v_\mu)u(v)=0$, and $\bar{u}_\lambda(v)(\qs-q\cdot v)u(v)=0$, we find a zero
total amplitude.

   The last example is the decay $\Sigma_b\to\Lambda_c\pi\ell\nu$,
whose amplitude at zero recoil is given by (see Eqs.~(4.46) and (4.47) of [1])
\be
\la\Lambda_c (v, s^\prime) \pi^d (q)|J_\mu^W|\Sigma^e_b (v, s)\ra
= i \frac{g_2}{\sqrt{2} f_\pi} \delta_{ed}C_{cb}\overline{u}(v, s^\prime)
[ E_{a\mu} + E_{b\mu} + E_{c\mu}] u (v, s).
\en 
Again, $E_{a\mu}$, $E_{b\mu}$ and $E_{c\mu}$ are the contributions
from Fig.~5(a), 5(b)
and 5(c) respectively:
\be
E_{a\mu} &=& \zeta(1)\frac{1}{-v\cdot q + M_{\Sigma_b}-M_{\Lambda_b}}
\gamma_\mu (1-\gamma_5) (\qs- q \cdot v ),  \\  
E_{b\mu} &=& - \frac{1}{3}\xi_1(1)\frac{1}{v\cdot q + M_{\Lambda_c}-M_{
\Sigma_c}}( \qs - q \cdot v ) [\gamma_\mu(1-\gamma_5) - 4 v_\mu\gamma_5 ], \\
E_{c\mu} &=& - 2 \xi_1(1)\frac{1}{v\cdot q + M_{\Lambda_c} -
M_{\Sigma^\ast_c}} [ -q_\mu + \frac{1}{3} (\qs - q \cdot
v) \gamma_\mu + \frac{1}{3} (\qs + 2q \cdot
v) v_\mu ] (1 + \gamma_5).
\en
Using the identity
\be
\bar{u}(v)\qs\gamma_\mu\gamma_5 u(v)=\,\bar{u}(v)(q\cdot v\gamma_\mu-\qs v_\mu)
\gamma_5 u(v)
\en
together with $\bar{u}(v)\gamma_5 u(v)=0$, and noting that $M_{\Sigma_b}-
M_{\Lambda_b}=M_{\Sigma_c}-M_{\Lambda_c}=M_{\Sigma^*_c}-M_{\Lambda_c}$ in
the heavy quark limit, we find again a vanishing total amplitude.
The equality of the mass differences noted above and the
relation $\zeta(1)=\xi_1(1)=1$ are crucial to obtaining a null amplitude. Both
features are very general and are essential to the general treatment given in
next section.

  We remark that all the above calculations can be greatly simplified
in the ``superfield" framework in which a pseudoscalar and a vector meson
field are combined into a meson superfield [2], and likewise a baryon 
superfield [8] for spin-${1\over 2}$ and spin-${3\over 2}$ sextet baryon 
fields. Feynman rules in terms of superfields become much simpler, and there 
are fewer Feynman diagrams to evaluate. Moreover, this method has the 
advantage that, as we shell see in the next section, the numerator is the 
same for all the diagrams with a fixed number of soft particles emitted
at zero recoil as long as the particles are emitted in the same sequence.
This feature enables us to generalize the analysis in this section to an
arbitrary number of light hadron emissions and to include the excited heavy
hadrons in the intermediate and final states, as elucidated in the next
section.

\vskip 0.6mm
\noindent{\bf III.~~Suppression of Light Meson Emissions at Zero Recoil}

    In this section we will establish within the superfield framework
that the amplitudes for $\bar{B}\to
D(D^*)+n\pi+\ell\nu$ with $n\geq 1$ vanish at zero recoil.
Here, a pion is used generically to denote any light hadron. We will 
proceed in several steps. First, we will only include the ground states
$\bar{B},~\bar{B}^*,~D,~D^*$ and Goldstone bosons in our discussion. Once we
have established the result for this simple system, we will generalize it to
include the excited heavy mesons in the intermediate and final states, and 
light particles other than the Goldstone bosons. Finally, we outline briefly
a similar analysis for the heavy baryons.

   The octet of Goldstone bosons is represented by the matrix
\be
M=\left(\matrix{ {\pi^0\over\sqrt{2}}+{\eta\over\sqrt{6}} & \pi^+ & K^+  \cr
\pi^- & -{\pi^0\over\sqrt{2}}+{\eta\over\sqrt{6}} & K^0  \cr
K^- & \bar{K}^0 & -\sqrt{2\over 3}\eta   \cr}\right).
\en
The ground states of the heavy mesons are the pseudoscalar $P(v)$ and vector
$P^*_\mu(v)$ with the quark content $Q\bar{q}$. The spin symmetry of the heavy
quark is incorporated automatically by the use of a superfield matrix, which
combines $P(v)$ and $P^*_\mu(v)$ [2]
\be
&& H^a_{Ii}(v)=\left\{ {1+v\!\!\!/ \over 2}[-P^a(v)\gamma_5+P^{a*}_\mu(v)
\gamma^\mu]\right\}_{Ii},   \\
&& \ov{H}^a=\gamma^0H^\dagger\gamma^0,
\en
where the indices $I$ and $i$ refer to the heavy quark and light antiquark,
respectively; the label $a$ indicates the SU(3) flavor of the light
antiquark. 

  The Lagrangian under our consideration is
\be
{\cal L} = {f_\pi^2\over 4}\,{\rm tr}\left(\partial_\mu \Sigma^\dagger
\partial^\mu\Sigma\right) +\sum_{Q=c,b}\bigg\{i{\rm tr}(H_Qv\cdot D\ov{H}_Q)+
g\,{\rm tr}(\ov{H}_QH_Q {\cal A}\!\!\!/ \gamma_5)\bigg\},
\en
where $f_\pi=94$ MeV, 
\be
\Sigma &=& \exp\left({2iM\over \sqrt{2}f_\pi}\right),   \\
D\ov{H} &=& (\partial_\mu+{\cal V}_\mu)\ov{H},
\en
and
\be
{\cal V}_\mu &=& {1\over 2}(\xi^\dagger\partial_\mu\xi+\xi\partial_\mu\xi^
\dagger)={1\over f_\pi^2}[M,~\partial_\mu M]+\cdots,   \\
{\cal A}_\mu &=& {i\over 2}(\xi^\dagger\partial_\mu\xi-\xi\partial_\mu\xi^
\dagger)=-{\sqrt{2}\over f_\pi}\partial_\mu M+\cdots,  \\
\xi &=& \Sigma^{1/2}.
\en
In (22) we have used a dimension ${3\over 2}$ field $H$ to absorb all the
heavy mass factors. The structure of the interaction vertices in Eq.(22)
shows that the light mesons are coupled to light quark's degrees of freedom
and the interactions are independent of the spin and flavor of the heavy 
quark. The heavy meson's propagator plays a crucial role in our discussion. 
It is given by
\be
\int d^4xe^{iq\cdot x}\la 0|TH^a_{Ii}(x)\ov{H}^b_{jJ}(0)|0\ra=-{i\over 
v\cdot q}\left({1+v\!\!\!/\over 2}\right)_{IJ}\left({1-v\!\!\!/ \over 2}
\right)_{ji}\delta_{ab}.
\en
It should be noted that the Dirac indices for the heavy quark $(IJ)$ and the 
light antiquark $(ji)$ decouple. The weak currents for $b\to c$ transitions in 
terms of the heavy meson fields are
\be
J_\mu^W=-C_{cb}\xi(v\cdot v')\,{\rm tr}\left[\ov{H}_c(v')\gamma_\mu(1-\gamma_5)
H_b(v)\right],
\en
where $C_{cb}$ is a QCD correction factor and $\xi(v\cdot v')$ is the 
Isgur-Wise function.

  Let us start with the diagram in which all the $n$ pions are emitted before
the weak vertex. We now move the pions one by one through the weak vertex
without changing their quantum numbers (momentum and SU(3) flavor), as 
depicted in Fig.~6. Because the initial and final heavy mesons must have 
the same velocity $v$, the weak 
vertex acts only on the heavy quarks, and the pions are emitted from the light
quarks, all the $(n+1)$ diagrams described above have a common numerator
(see Fig.~7)
\be
N_n=-g^n{\rm tr}\left[\ov{H}'\gamma_\mu(1-\gamma_5)
H{\cal A}\!\!\!/ _1\gamma_5{1-v\!\!\!/ \over 2}{\cal A}\!\!\!/_2
\gamma_5\cdots{1-v\!\!\!/ \over 2}{\cal A}\!\!\!/ _n\gamma_5\right].
\en
For a single pion emission ${\cal A}_i^\mu\sim {1\over 2}\lambda^{a_i}q_i^\mu$.
The wave functions for the initial and final mesons are denoted by
$H$ and $\ov{H}'$, respectively; for a pseudoscalar meson $H=-{(1+v\!\!\!/)
\over 2}\gamma_5$, and for a vector meson $H={(1+v\!\!\!/)\over 2}
\ep\!\!\!/$, where $\ep^\mu$ is a polarization vector. We have not 
explicitly showed the SU(3) flavor wave functions.
We have ignored the QCD factor $C_{cb}$ and the
Isgur-Wise function is unity. The amplitude for the $(n+1)$ diagrams is given 
by
\be
M(q_1,\cdots,q_n)=N_n A(q_1,\cdots,q_n),
\en
with
\be
A(q_1,\cdots,q_n) &=& \sum^n_{i=0}(-1)^{n-i}{1\over v\cdot q_1}\cdot{1\over
v\cdot (q_1+q_2)}\cdots{1\over v\cdot(q_1+q_2+\cdots+q_{n-i})}  \\ 
&& \cdot{1\over v\cdot(q_{n-i+1}+\cdots+q_n)}\cdot{1\over v\cdot(q_{n-i+2}+
\cdots+q_n)}\cdots{1\over v\cdot(q_{n-1}+q_n)}\cdot {1\over v\cdot q_n}. \non
\en
For $1\leq i\leq n-1$, the two propagators adjacent to the weak vertex can be 
rearranged as follows:
\be
&& {1\over v\cdot(q_1+q_2+\cdots+q_{n-i})}\cdot{1\over v\cdot(q_{n-i+1}+q_{
n-i+2}+\cdots+q_{n})}   \\
&& =\left[{1\over v\cdot(q_1+q_2+\cdots+q_{n-i})}+{1\over v\cdot(q_{n-i+1}+q_{
n-i+2}+\cdots+q_{n})}\right]\cdot{1\over v\cdot(q_1+q_2+\cdots+q_n)}. \non
\en
The summation over $i=1$ to $i=n$ is now broken into two series. We note that
the two series almost cancel each other except for the first term of the first
series and the last term of the second series. These two terms in turn
cancel exactly the $i=0$ and $i=n$ terms in the original series. Therefore,
\be
A(q_1,\cdots,q_n)=0.
\en
This is the central result of our paper. In what follows we make a series of
generalizations:
 
   (A).~~In writing down the amplitudes, we have implicitly assumed that a 
single particle is emitted at each vertex, but this is not necessary. If more 
than one particles are emitted at a particular vertex, e.g., in seagull
diagrams, the momentum $q_i$ is 
simply the sum of the total momenta of emitted particles. The proof still
goes through. Thus, the multiparticle vertices  in ${\cal A}_\mu$ and 
${\cal V}_\mu$ of Eqs.(25-27) can be included. Another class of diagrams can 
be incorporated in a similar fashion: a single pion can turn into 
multipions via the nonlinear interactions among the Goldstone bosons. 

   (B).~~ It is easy to include emissions of light particles other than the
Goldstone bosons. All it requires is that their interactions with the heavy
hadrons be independent of the flavors and spin of the heavy quarks. Then the
numerators for all the diagrams in Fig.~6 will be the same, and the same
result will follow.

   (C).~~We now consider a somewhat nontrivial generalization to include 
excited states in the intermediate states and final states. As in all previous
cases, the interactions between the light particles and heavy mesons are
independent of heavy quark's spin and flavor. At the point of no recoil,
the weak vertex is nonzero only if both sides belong to the same 
supermultiplet. (States in a supermultiplet differ only in how the heavy 
quark spin is combined with light antiquark states of a definite angular 
momentum
and parity.) Moreover, there is only one Isgur-Wise function contributing 
and it has the value unity; the weak vertex has the standard $(V-A)$ 
combination acting on the 
heavy quarks. Thus, the numerator factor for the diagrams similar to
Fig.~6 is of the same form as in (30) with a different and possibly more 
complicated matrix ${\cal A}$ at each vertex. The excited states and the 
ground 
states in general have mass differences of order 1 due to excitations of the
light quark system; they may have an order 1 decay widths. All these mass
differences and decay widths respect heavy quark symmetry. (In contrast,
the mass differences and decay widths within a supermultiplet are of order
$1/m_Q$ or smaller and they violate heavy quark symmetry.) To account for
these we assign a complex mass $M_i$ to the intermediate state which 
follows the emission of the $i$th pion (see Fig.~8).
More precisely, we will use the 
notation $M_i(Q)$ if the intermediate state appears before the weak vertex, 
and $M_i(Q')$ if it appears after the weak vertex. Let us define
\be
\Delta_i &=& M(Q)-M_i(Q)  \non \\
         & =& M(Q')-M_i(Q'),~~~~i=1,2,\cdots,n,       
\en
where $M(Q)$ is the mass of the initial state, and $M_n(Q')$ is the mass of the
final state. The masses $M(Q')$ and $M_n(Q)$ are, respectively, those of the
corresponding states with the heavy quarks $Q$ and $Q'$ interchanged. The two
forms of (35) follow from the heavy flavor independence of the mass
differences. Consider two examples. When all the pions are emitted after 
the weak vertex, the intermediate state after the weak vertex has the 
same quantum numbers as the initial state except the heavy quark $Q$ is 
replaced by $Q'$. According to (35), the mass of this intermediate state 
is denoted by $M(Q')$. Likewise, when all the pions are emitted before the 
weak vertex, the intermediate state before
the weak vertex has the same quantum numbers as the final state except its 
heavy quark and its mass is denoted by $M_n(Q)$. 
The amplitude corresponding to (32) becomes
\be
A'(q_1,\cdots,q_n) &=& \sum^n_{i=0}(-1)^{n-i}{1\over v\cdot q_1-\Delta_1}
\cdot{1\over v\cdot (q_1+q_2)-\Delta_2}\cdots{1\over v\cdot(q_1+q_2+\cdots+
q_{n-i})-\Delta_{n-i}}  \non \\ 
&& \cdot{1\over v\cdot(q_{n-i+1}+\cdots+q_n)+\Delta_{n-i}-\Delta_n}
\cdots{1\over v\cdot(q_{n-1}+q_n)+\Delta_{n-2}-\Delta_n} \non \\
&& \cdot {1\over v\cdot q_n+\Delta_{n-1}-\Delta_n}. 
\en
One may repeat the same analysis following steps similar to (33), or one
may simply observe that (36) can be reproduced from (32) by the 
substitutions
\be
&& v\cdot q_1 \to v\cdot q_1-\Delta_1,  \non \\
&& v\cdot q_2 \to v\cdot q_2+\Delta_1-\Delta_2,  \non \\
&& \cdots   \\
&& v\cdot q_i \to v\cdot q_i+\Delta_{i-1}-\Delta_i,  \non \\
&&\cdots  \non \\
&& v\cdot q_n \to v\cdot q_n+\Delta_{n-1}-\Delta_n.  \non 
\en
In the rest frame of $v$, we have $v\cdot q_i=q_i^0$, and the substitutions
become simple shifts in $q_i^0$'s. Since the result (34) holds for
arbitrary values of $v$, and $q_1,\cdots,q_n$, we conclude that
\be
A'(q_1,\cdots,q_n)= 0.
\en

   (D).~~We now consider how to include closed 
loops. A closed loop can be produced by setting $q_i=-q_j$ and supplying the 
appropriate propagator for a light particle of momentum $q_i$ and 
finally integrating over the loop momentum $q_i$. Since the amplitude vanishes
for arbitrary $q_i$ and $q_j$, this procedure will not alter the conclusion 
(34).  There is only a minor complication when the closed loops form 
self energy parts of the external legs. In these cases, the sum of the
momenta $q_i+\cdots+q_l$ in the loop(s) vanishes and some of the 
propagators become singular. We can regulate these singularities by supplying
temporarily small, but nonzero, mass differences to these lines similar to the
case of excited states. This procedure is equivalent to lifting the
external legs slightly off the mass shell.

   (E).~~It is straightforward to generalize our discussion to heavy baryons. 
Introducing the superfields
\be
S^\mu &=& B_6^{*\mu}-{1\over\sqrt{3}}(\gamma^\mu+v^\mu)\gamma_5 B_6,  \non \\
\bar{S}^\mu &=& \bar{B}_6^{*\mu}+{1\over\sqrt{3}}\bar{B}_6\gamma_5
(\gamma^\mu+v^\mu),    \\  
T &=& B_{\bar{3}},  \non
\en
where $B_{\bar{3}},~B_6$ and $B^*_6$ are baryon component fields for 
spin-${1\over 2}$ antitriplet, sextet, and spin-${3\over 2}$ sextet baryon 
states in SU(3) flavor representations, respectively (see ref.[1] for 
details).  The relevant chiral 
Lagrangian and weak Lagrangian are [8,9]
\be
{\cal L}_B &=&-i\tr(\bar{S}^\mu v\cdot DS_\mu)+{i\over 2}\tr(\overline
{T}v\cdot DT)+\Delta\,\tr(\bar{S}^\mu S_\mu)  \non \\
&&+i{3\over 2}\,g_1\epsilon_{\mu\alpha\beta\nu}\tr(\bar{S}^\mu v^\alpha{\cal A}
^\beta S^\nu)-\sqrt{3}\,g_2\tr(\bar{S}^\mu{\cal A}_\mu T)+h.c., \\
{\cal L}_W &=&C_{ji}\bigg\{\overline{S}'^\lambda(v')\gamma_\mu(1-\gamma_5)S^
\kappa(v)[-g_{\lambda\kappa}\xi_1(v\cdot v')+v_\lambda v'_\kappa\xi_2(v\cdot 
v')] \non\\
&& +\overline{T}'(v')\gamma_\mu(1-\gamma_5)T(v)\zeta(v\cdot v')\bigg\},
\en
where $\Delta=m_S-m_T$ is the mass splitting between the sextet and 
antitriplet baryon multiplets.
As mentioned before, $\zeta(1)=\xi_1(1)=1$ and $\xi_2(1)$ drops out in 
the amplitude. From the examples of Eqs.(40) and (41) it is clear that 
because of the spin and flavor symmetry of the heavy quarks, it is
also true in the heavy baryon sector that (i) the coupling constants and
structure of interactions between light particles and heavy baryons are the 
same for baryons containing a different heavy quark, and (ii) the numerator
for the matrix element of emission of $n$ light particles is the
same regardless of the number of particles emitted before and after the
weak vertex as long as the $n$ particles are emitted in the same sequence. 
This is because the Dirac matrices only appear at the weak vertex which
are coupled to the heavy quark indices. Moreover, the only contributing
Isgur-Wise function has the universal value unity.
For example, the numerator for the amplitude of 
$S\to S'+n\pi+\ell\nu$ is of the form $\bar{S}'^\beta\gamma_\mu(1-\gamma_5)
\cdots S_\alpha$, where the expression represented by the ellipses $\cdots$ 
depends on the intermediate states involved but are free of Dirac matrices.
The light quarks in a heavy baryon
behave as a whole like a Bose system and the vertices for light particle
emissions involve the same string of vectorial indices for all
the diagrams in Fig.~8 with heavy mesons replaced by heavy baryons.
Therefore, these diagrams will lead to a partial amplitude 
proportional to $A(q_1,\cdots,q_n)$ of Eq.(34). Thus, light particle 
emissions are suppressed at zero recoil.

\vskip 0.6mm
\noindent{\bf IV.~~Photon Emissions}

   As pointed out in the Introduction, photon emissions are interesting because
the heavy quark's electromagnetic interactions violate heavy quark's flavor 
symmetry , and the coupling strengths for pion and photon emission are
comparable (the former is given by ${q\over f_\pi}g$ with $q$ being the
pion momentum, and the latter is given by $e$.) It is therefore important
to investigate whether light particle emissions are still suppressed at 
the point of no recoil in the presence of electromagnetic interactions. 

    However, photons come from several different sources. Some of these photons
preserve heavy quark symmetry. Among these are the photons emitted by
the usual convection currents and magnetic moment (Pauli terms) couplings
of the light quarks in the heavy hadrons. These photons
can be included in the category of other light particles discussed in the last
section. Photons can also be emitted by the charged light particles
themselves, such as pions and $\rho$ mesons. These interactions belong to
another category already treated in the last section in which many particles 
are emitted at a single vertex. Finally, photons can be emitted from
the heavy quarks. A heavy quark's magnetic moment is of order
$1/m_Q$ and hence it can be neglected in our present discussion. Thus, we 
only have to deal with the photons emitted by the heavy quark's 
convection currents.

    We will show in this section that the suppression of light hadron
emissions at the point of no recoil still presisits when photon emissions from
the heavy quarks are included. The reason is that a heavy quark's convection
current is effectively an identity operator provided the emitted
photon's momentum is much smaller than the heavy quark mass $m_Q$. In terms of
Feynman diagrams, the above statement translates into a factorization of the 
photon emission amplitude and the light hadron emission amplitude. At the 
end of this section, we will conclude that photon emissions accompanied
by any light hadron emission is forbidden at the point of no recoil.
The only photons emitted are those from the heavy quarks and the charged 
lepton and are to accompany the basic processes $\bar{B}\to D(D^*)+\ell\nu$.

    We should emphasize the distinction between the factorization of photon
emissions and the light hadron emissions here and the well-known result in
soft photon physics. The photons emitted here are soft compared with the 
heavy quark mass $m_Q$, but they are not soft compared with the other light
hadrons emitted. Consequently, we have to include photons emitted
from the interior of a Feynman diagram, while the really soft photons can
only be emitted from the external legs.

   We begin by considering the part of the amplitude in which a photon
of momentum $k$ and $n_1$ light hadrons of momenta $q_1,\cdots,q_{n_1}$
are emitted from the initial heavy quark (see Fig.~9).
The photon can be emitted anywhere
on the initial heay quark line, and the light hadrons with momenta $q_1,
\cdots q_{n_1}$ are emitted from the initial heavy quark in a specified
sequence. This partial amplitude is proportional to
\be
&& A(k,q_1,\cdots,q_{n_1})=(-1)^{n_1+1}e_Q(v\cdot\varepsilon)  \\
&& \times\sum^n_{i=0}{1\over v\cdot q_1}\cdot{1\over
v\cdot (q_1+q_2)}\cdots{1\over v\cdot(q_1+q_2+\cdots+q_{i})}\cdot{1\over
v\cdot(q_1+\cdots+q_i+k)}   \non \\ 
&& \cdot{1\over v\cdot(q_1+q_2+\cdots+q_i+q_{i+1}+k)}\cdots{1\over v\cdot(q_1
+q_2+\cdots+q_{n_1}+k)},   \non
\en
where $e_Q$ is the electric charge of the initial quark and $\varepsilon$ is 
the photon's polarization vector. Adding up the terms one by one sequentially 
in the order of increasing $i$, we find the factorized form
\be
A(k,q_1,\cdots,q_{n_1})=\left(-e_Q{v\cdot\ep\over v\cdot k}\right)\bigg[(-1)
^{n_1}{1\over v\cdot q_1}\cdot{1\over v\cdot(q_1+q_2)}   
\cdots{1\over v\cdot(q_1+q_2+\cdots+q_{n_1})}\bigg].
\en

 The above result can be stated in words: insertion of a photon anywhere on 
the initial heavy quark line gives rise to a factorized amplitude as a product
 of the one photon emission amplitude and the amplitude of multiparticle
emission. We can repeat the procedure by inserting the second photon 
everywhere, etc. We then find the amplitude for $l_1$ photons emitted in 
addition to $n_1$ light hadrons emitted from the initial heavy quark to be
given by
\be
A^i(k_1,\cdots,k_{l_1},q_1,\cdots\,q_{n_1}) &=& \left(-e_Q{v\cdot\ep_1\over
v\cdot k_1}\right)\cdots\left(-e_Q{v\cdot\ep_{l_1}\over v\cdot k_{l_1}}\right) 
\\
&\times & \left[(-1)^{n_1}{1\over v\cdot q_1}\cdot{1\over v\cdot (q_1+q_2)}
\cdots{1\over v\cdot(q_1+\cdots+q_{n_1})}\right].  \non
\en
A similar result holds for emissions of $l_2$ photons and $n_2$ light hadrons
from the final heavy quark of electric charge $e_{Q'}$ (Fig.~10):
\be
&& A^i(k_{l_1+1},\cdots,k_{l_1+l_2},q_{n_1+1},\cdots\,q_{n_1+n_2}) = \left(e_
{Q'}{v\cdot\ep_{l_1+1}\over v\cdot k_{l_1+1}}\right)\cdots\left(e_{Q'}{v\cdot
\ep_{l_1+l_2}\over v\cdot k_{l_1+l_2}}\right) \non\\
&& \times\left[{1\over v\cdot(q_{n_1+1}+\cdots+q_{n_1+n_2})}\cdot
{1\over v\cdot(q_{n_1+2}+\cdots+q_{n_1+n_2})}\cdots{1\over v\cdot
q_{n_1+n_2}}\right].
\en
We now combine the two results (44) and (45) by keeping fixed both the
$n_1$ and $n_2$ particles emitted by the two heavy quarks, respectively, and
the total number of photons emitted, $l_1+l_2=l$, varying $l_1$ and $l_2$
and permuting the photons in all possible ways. The result is
\be
&& A(k_1,\cdots,k_l,q_1,\cdots,q_{n_1},q_{n_1+1},\cdots,q_{n_1+n_2}) \non \\
&& =\prod_{i=
1}^l(-e_Q+e_{Q'})\,{v\cdot\ep_i\over v\cdot k_i} \cdot\bigg[(-1)^{n_1}{1\over
v\cdot q_1} \cdot{1\over v\cdot(q_1+q_2)}\cdots {1\over v\cdot(q_1+\cdots+q_{
n_1})}  \non \\
&& \cdot{1\over v\cdot(q_{n_1+1}+\cdots+q_{n_1+n_2})}   
\cdot{1\over v\cdot(q_{n_1+2}+\cdots+q_{n_1+n_2})}\cdots{1\over v\cdot
q_{n_1+n_2}}\bigg].
\en
The result (46) holds for any choice of $n_1$ and $n_2$. Adding up all 
contributiuons with the fixed sum $n_1+n_2=n$ and keeping the sequence of 
light hadrons unchanged as they move across the weak vertex, we find the
amplitude for emission of $l$ photons and $n$ light hadrons in a definite
sequence is
\be
A(k_1,\cdots,k_l,q_1,\cdots,q_n) &=& \prod^l_{i=1}(-e_Q+e_{Q'})\,{v\cdot\ep_i
\over v\cdot k_i}\,A(q_1,\cdots,q_n)   \non \\
&=& 0,
\en
where $A(q_1,\cdots,q_n)$ is given by (32) and (34).

   To include the excited states in the above discussion, one proceeds
exactly the same as the discussion on particle emission. The same conclusion 
(47) holds. To be precise, we should have included photon emissions from the 
charged lepton in the above amplitude. Since these contributions are
already clearly in the factorized form, they will not affect the
conclusion above. Eq.(47) states that at the point of no recoil, suppression 
of light hadron emissions still persists in the presence of photon
emissions from the heavy quarks.

   For completeness, we give the results for photon emissions which decorate
the basic processes $\bar{B}\to D(D^*)+\ell\nu$. To arrive at a gauge
invariant result, it is necessary to include photon emissions from the 
charged lepton. In addition, to obtain a simple closed form, we will
assume that photons are soft compared with the charged lepton momentum.
Under these conditions, the leading contributions from the charged lepton
are due to its convection current. The treatment of summing up these photon 
emissions is well-known [10]. The result at zero recoil involves the basic
amplitude for a single photon emission
\be
A(k)=(-e_Q+e_{Q'})\,{v\cdot\ep\over v\cdot k}+e_\ell\,{v_\ell\cdot\ep\over
v_\ell\cdot k},
\en
where $e_\ell$ and $v_\ell$ are the electric charge and velocity of the charged
lepton. Eq.(48) is gauge invariant as a result of the charge conservation
$-e_Q+e_{Q'}+e_\ell=0$. Contributions from arbitrary number of virtual photons
supply a multiplicative factor to the amplitude for $\bar{B}\to D(D^*)+\ell
\nu$:
\be
V &=& e^X,   
\en
with
\be
X &=& {1\over 2}e^2_\ell\int{d^4k\over (2\pi)^4}\,{i\over k^2+i\epsilon}\left(
{v\over v\cdot k}-{v_\ell\over v_\ell\cdot k}\right)^2,~~~~ \lambda<|\vec{k}|
<\Lambda.
\en
The correction to the rate due to virtual photons is then given by the
factor
\be
&&|V|^2=e^{2{\rm Re}X}=\left({\lambda\over\Lambda}\right)^A,   \\
&& A=-{2\alpha\over \pi}\left\{1-{v\cdot v_\ell\over\sqrt{(v\cdot v_\ell)^2-1}}
\ln[v\cdot v_\ell+\sqrt{(v\cdot v_\ell)^2-1}]\right\}.
\en
Summing up the contributions from all real photons leads to the multiplicative
factor to the rate
\be
{\cal R}=\left({E\over\lambda}\right)^AF({E\over E_T},A),
\en
where
\be
F(x,A)={1\over \pi}\int^\infty_{-\infty}{du\over u}{\sin}u\exp\left[A\int^x
_0{dk\over k}\left(e^{iku}-1\right)\right].
\en
The energy parameters $E$ and $E_T$ are introduced in the constraints for
individual photon energy and the total energy carried by the photons:
\be
&& \lambda<k_i<E,   \\
&& \sum_i k_i< E_T.
\en
The virtual and real photons together produce a correction factor to the
rate for $\bar{B}\to D(D^*)+\ell\nu$:
\be
e^{2{\rm Re}X}\cdot{\cal R}=\left({E\over\Lambda}\right)^AF({E\over E_T},A),
\en
where $\Lambda$ is a factorization scale for separating soft and hard 
photons, and $E$
is the energy below which a photon cannot be detected. 
The $\Lambda$ dependence in (57) is canceled by the hard photon corrections 
to the cross sections for $\bar{B}\to D(D^*)+\ell\nu$.
A judicious choice of $\Lambda$ will minimize the effects of
hard photons, and (57) will be the dominant radiative correction factor. For
instance, this occurs when $A\ln (W/\Lambda) <<1$, where $W$ is the typical 
energy in the process. For $B$ decays at zero recoil, there is a wide 
latitude for such a choice of $\Lambda$. 

   Identical discussions apply to the CKM favored semileptonic decays
of heavy baryons. Again, the photons are
the only light particles that can be emitted in this limit; and
they are to accompany the simple processes $S_b\to S_c+\ell\nu$, $T_b\to
T_c+\ell\nu$. The photon emissions are described by the same 
multiplicative factors given by (57).

\vskip 0.6mm
\noindent{\bf V.~~Connection between Formal Argument and Diagrammatic
Analysis}

   An important and interesting result such as the one discussed in this paper
deserves, on the one hand, a derivation from some general principle, and, on 
the other hand, a deeper understanding how it actually works in a concrete
framework. The original argument in [4,5] is of the first kind, while our
discussion of the last two sections is of the second kind. In this section,
we offer a proof of the key result (34) which provides a link between the
general argument and the diagrammtic analysis. In doing so we have discovered
that the general result does not require the full power of the formal
argument.  It is hoped that our work will stimulate more research on this
subject. 

   The key result (34) is derived without reference to the details of the
vertices ${\cal A}\!\!\!/ (q_i)$ in (30) except that they are independent of 
heavy quark's spin and flavor. For all purposes,  they can be regarded as
external sources. We are motivated to consider the vertex function of the
weak current
\be
{\cal J}_\mu(v\cdot v'|K)\equiv\la
H_c(v')|J_\mu^W(0)|H_b(v)\ra=-\xi(v\cdot v'|K){\rm tr}[\ov{H}_c(v')
\gamma_\mu(1-\gamma_5)H_b(v)],
\en
where $H_Q(v)$ denotes the wave function of a heavy meson with quark content
$Q\bar{q}$. Its explicit form is ${1+v\!\!\!/\over 2}\ep\!\!\!/$ for a 
vector meson and ${1+v\!\!\!/\over 2}(-\gamma_5)$ for a pseudoscalar meson.
The argument $K$ in the Isgur-Wise function emphasizes the fact that
$\xi$ is a functional of the external source $K(x)$ which appears in the
interaction (see Fig.~11)
\be
{\cal L}_K=\sum_{Q=c,b}{\rm tr}(\ov{H}_QH_QK),
\en
where
\be
K(x)=\sum_i K_i(x)t_i,
\en
and $t_i$ is possibly a combination of Dirac and SU(3) flavor matrices. The
external source respects heavy quark symmetry. We will further assume that
its Fourier component $K(q)$ carries momentum much smaller than $m_Q$. The
standard argument still applies to obtain
\be
\xi(v\cdot v'=1|K)=1,
\en
which states that the external source does not alter the state of a heavy 
quark, and the weak current at the point of no recoil can only convert a 
$\bar{B}$ meson into a $D$ or $D^*$ meson. Now, the right-hand side of (61) 
is independent of the external source. Consequently, its functional
derivatives evaluated at zero source must vanish:
\be
{\delta^n\over \delta K(q_1)\delta K(q_2)\cdots\delta K(q_n)}\,\xi(v\cdot
v'=1|K)\bigg |_{K=0}=0.
\en
Or equivalently,
\be
{\delta^n\over \delta K(q_1)\delta K(q_2)\cdots\delta K(q_n)}\,{\cal J}_\mu
(v\cdot v'=1|K)\bigg |_{K=0}=0.
\en
A simple evaluation gives
\be
{\delta^n\over \delta K(q_1)\delta K(q_2)\cdots\delta K(q_n)}\,{\cal J}_\mu
(v\cdot v'=1|K)\bigg |_{K=0} &=& N_n'A(q_1,\cdots,q_n)  \non \\
&+& {\rm all~permutations~in~}(1,\cdots,n),
\en
where $N_n'$ is very similar to $N_n$ in (30):
\be
N_n'=-{\rm tr}\left[\ov{H}_c\gamma_\mu(1-\gamma_5)
H_bt_1{1-v\!\!\!/ \over 2}t_2\cdots{1-v\!\!\!/ \over 2}t_n\right],
\en
and $A(q_1,\cdots,q_n)$ is the same as (32). Eqs.(61) and (62) 
together establish the connection between the formal and diagrammatic 
appraches. However, we find it interesting that our explicit calculation
in Section 3 shows that $A(q_1,\cdots,q_n)=0$ all by itself without 
having to combine with similar contributions from permutations. Does this
fact imply that there is more information in (34) than what we can extract
so far ? On the other hand, the numerator $N_n'$ given by (65) is not symmetric
with respect to the light particles. It contains noncommuting Dirac and/or
SU(3) flavor matrices, and each source may refer to different species of
particles. Were it essential to include all the $n!$ sets of diagrams, the 
analysis in Section 3 would have been much more complicated.
   The utility of employing an external source allows us to, effectively, 
deal with all the multiparticle final states at once. The single equation (61)
contains all the physics needed to derive the general result (62).

\vskip 0.6mm
\noindent{\bf VI.~~Discussions and Conclusions}

   In this section we discuss our results and their implications. First of 
all, let us recapitulate what we have accomplished in this work. We have 
established the result of refs.[4,5] by a diagrammatic analysis, and have 
generalized it to include photon emissions. The diagrammatic approach is
complementary to the formal arguments of refs.[4,5]. The two approaches are 
united by considering the Isgur-Wise function in the presence of an
external source which is independent of heavy quark's spin and flavor.

   Thanks to heavy quark symmetry, it is possible to make a precise statement
on some properties of systems containing a heavy quark. The subject
investigated in this paper is such an example. The conclusion of our work is 
the following exact statement:  For CKM favored semileptonic
decays of a $\bar{B}$ meson or a bottom baryon, the only light particles that 
can be emitted at
the point of no recoil are photons emitted from the heavy quarks and
the charged lepton. These photons are to accompany the basic decay $\bar{B}
\to D(D^*)\ell\nu$.  The corrections to the above statement are of
order $1/m_Q^2$ in the decay rates. Similar statements apply to the heavy
baryons $S_b$ and $T_b$.

  In the so-called Shifman-Voloshin limit [4], $(m_Q+m_{Q'})\Lambda_{\rm QCD}
<<(m_Q-m_{Q'})^2<<(m_Q+m_{Q'})^2$, the variable $v.v'= 1+{\cal O}
[(m_Q-m_{Q'})/(m_Q+m_{Q'})]^2$, so light particle emissions are severely
suppressed. The decays $\bar{B}\to D(D^*)+\ell\nu$ are not far from this limit.
Therefore, we do not expect $B$ decays involving additional pions or 
excited states of $D$ and $D^*$ to have significant rates. 

  The conclusion of this paper has at least another important practical
application. The fundamental parameter $V_{cb}$ of the Standard Model is 
directly related to the differential decay rate of $\bar{B}\to D^* +\ell\nu$
at $v=v'$. Since the Isgur-Wise function has the value unity at this point, 
$V_{cb}$ is the only unknown in this decay rate. At CESR the $B$ and $\bar{B}$
mesons are produced at rest, so the decay rate for $\bar{B}\to D +\ell\nu$ at 
$v=v'$ is vanishingly small compared with the $D^*$ mode due to conservation 
of angular momentum\footnote{The ratio for the two rates near zero recoil 
is $\Gamma_D:\Gamma
_{D^*}=v_D^2:12\left({m_b-m_c\over m_b+m_c}\right)^2$, where $v_D$ is the 
velocity of the $D$ meson.}.
According to the Luke's theorem, the corrections to the 
Isgur-Wise function is of order $1/m_Q^2$ while corrections due to light 
particle emissions and excited charmed meson states are also of the same 
order. It appears 
that using $\bar{B}\to D^* +\ell\nu$ at this
kinematic point to measure $V_{cb}$ is possibly the cleanest way.

In this paper we have
restricted ourselves to emissions of light particles with momenta small
compared with $m_Q$. This is necessary in order for us to apply heavy quark
symmetry. In principle, we can ensure that we are in this kinematic region for 
decays of $B$ mesons at rest by selecting lepton pairs with the total 
lepton energy to be close to $m_B-m_{D^*}$ [11].

\vskip 2.0cm
\centerline{\bf ACKNOWLEDGMENTS}

T.M.Y.'s work is supported in part by the
National Science Foundation.  This research is supported in part by the
National Science Council of ROC under Contract Nos.  NSC85-2112-M001-010,
NSC85-2112-M001-021.

\pagebreak

\vskip 1.5 cm
\centerline{\bf REFERENCES}
\vskip 0.3 cm
\begin{enumerate}

\item T.M. Yan, H.Y. Cheng, C.Y. Cheung, G.L. Lin, Y.C. Lin, and H.L.
Yu, {\sl Phys. Rev.} {\bf D46}, 1148 (1992); see also T.M. Yan, {\sl Chin. J.
Phys.} (Taipei) {\bf 30}, 509 (1992).

\item M.B. Wise, {\sl Phys. Rev.} {\bf D45}, R2188 (1992).

\item G. Burdman and J. Donoghue, {\sl Phys. Lett.} {\bf B280}, 287
(1992).

\item M.B. Voloshin and M.A. Shifman, {\sl Yad. Fiz.} {\bf 47}, 801 (1988)
[{\sl Sov. J. Nucl. Phys.} {\bf 47}, 511 (1988)].

\item C.G. Boyd, B. Grinstein, and A.V. Manohar, UCSD/PTH95-19 
[hep-ph/9511233].

\item M.E. Luke, \pl {\bf B252}, 447 (1990).

\item H.Y. Cheng, C.Y. Cheung, W. Dimm, G.L. Lin, Y.C. Lin, T.M. Yan, and 
H.L. Yu, {\sl Phys. Rev.} {\bf D48}, 3204 (1993).

\item P. Cho, \pl {\bf B285}, 145 (1992); \np {\bf B396}, 183 (1993).

\item H.Y. Cheng, C.Y. Cheung, G.L. Lin, Y.C. Lin, T.M. Yan, and H.L. Yu,
{\sl Phys. Rev.} {\bf D49}, 5857 (1994).

\item D.Yennie, S. Frautschi, and H. Suura, {\sl Ann. Phys.} {\bf 13}, 379
(1961). A concise discussion can be found in M. 
Peskin and D. Schroeder, {\it Introduction to Quantum
Field Theory} (Addison-Wesley Publishing
Company, NY, 1995), and S. Weinberg, {\it The Quantum Theory of Fields I}
(Cambridge University Press, NY, 1995).

\item We would like to thank Prof. Persis Drell for a useful conversation on
how CLEO measures the missing momentum and energy of neutrinos.

\end{enumerate}
\pagebreak

\centerline{\bf FIGURE CAPTIONS}
\vskip 1.0cm
\begin{description}

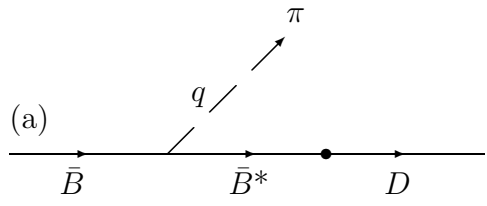
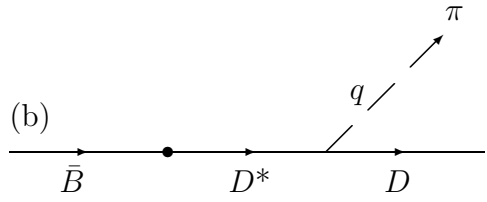
\item[Fig.~1] Feynman diagrams contributing to the decay $\bar{B} \rightarrow 
D +\pi + \ell \nu$.  The weak vertex is represented by a black dot.
The pion momentum is $q$.
 
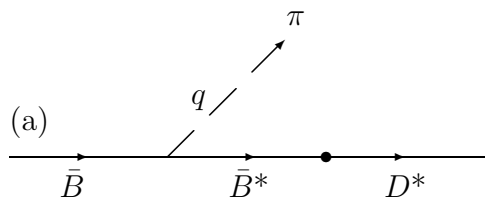
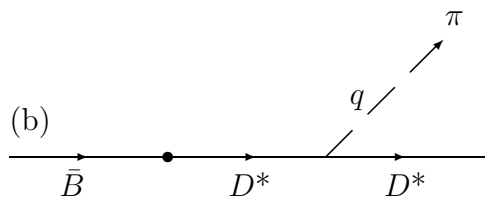
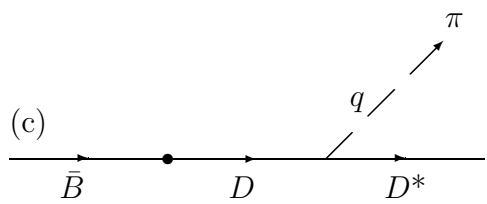
\item[Fig.~2] Feynman diagrams contributing to the decay $\bar{B} \rightarrow
D^\ast + \pi + \ell \nu$. The weak vertex is represented by a black dot.
The pion momentum is $q$.
 
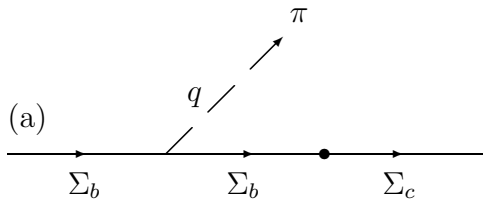
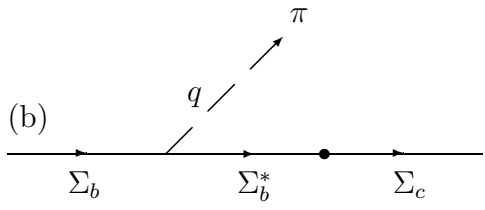
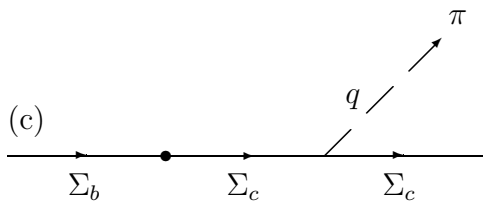
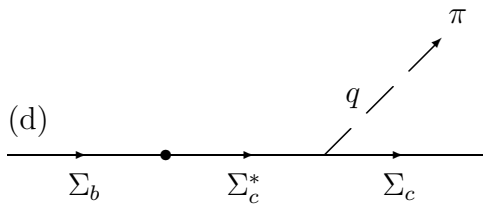
\item[Fig.~3] Feynman diagrams contributing to the decay $\Sigma_b
\rightarrow \Sigma_c + \pi + \ell \nu$.  The weak vertex is represented by 
a black dot. The pion momentum is $q$.
 
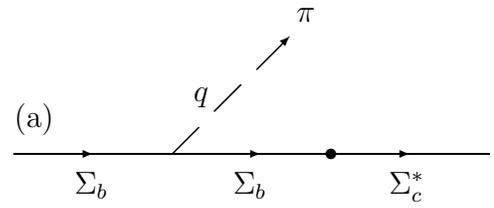
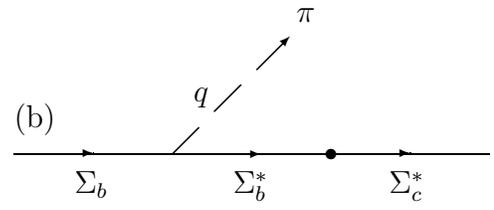
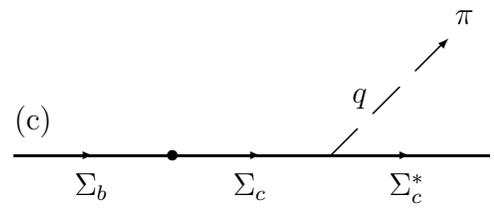
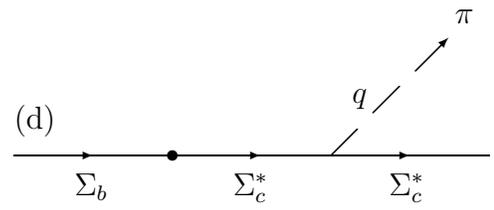
\item[Fig.~4] Feynman diagrams contributing to the decay $\Sigma_b
\rightarrow \Sigma^\ast_c + \pi + \ell \nu$.  The weak vertex is 
represented by a black dot. The pion momentum is $q$.
 
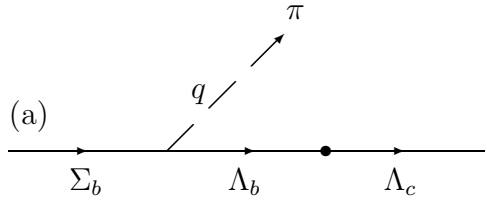
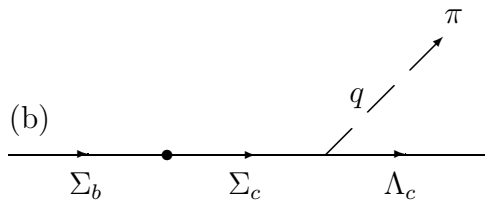
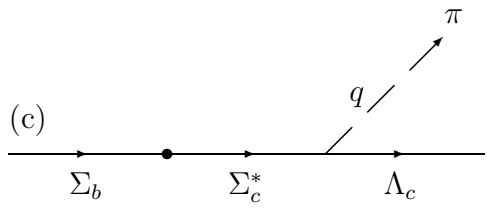
\item[Fig.~5] Feynman diagrams contributing to the decay $\Sigma_b
\rightarrow \Lambda_c + \pi + \ell \nu$.  The weak vertex is represented 
by a black dot. The pion momentum is $q$.

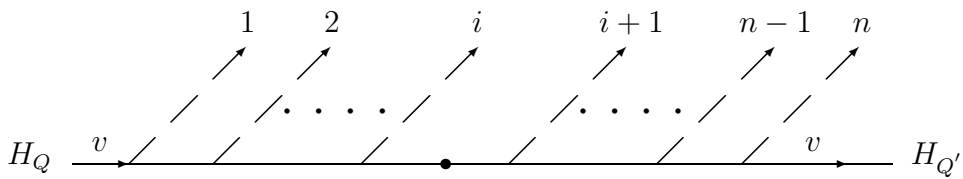
\item[Fig.~6] A typical Feynman diagram contributing to $H_Q\to H_{Q'} +\pi_1
+\pi_2 +\cdots+\pi_n + \ell + \nu$. The black dot denotes the weak vertex.

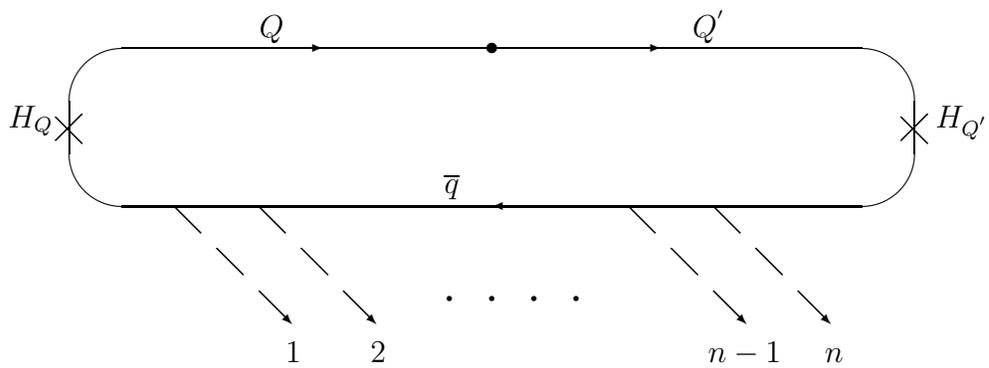
\item[Fig.~7] Graphic representation of the numerator of the amplitudes for 
all diagrams of the type in Fig.~6 with the pion sequence unchanged as the 
index $i$ varies from 0 to $n$. 

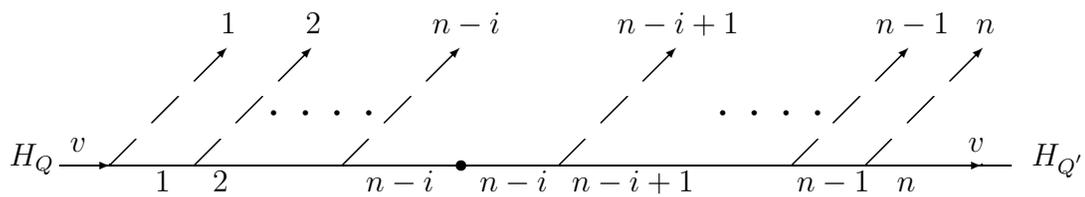
\item[Fig.~8] A typical Feynman diagram contributing to $n$ pion emissions
with excited heavy mesons as intermediate and final states. At the point of 
no recoil, the two states next to the weak vertex (the black dot) must have
the same light quark quantum numbers but with a different heavy quark.

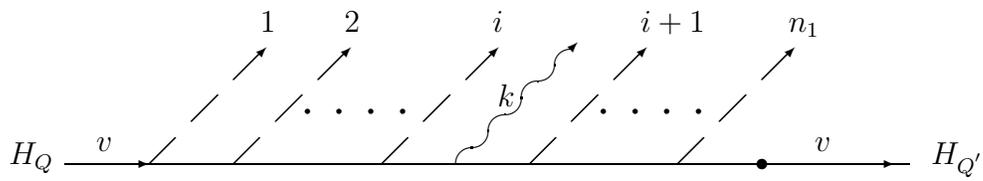
\item[Fig.~9] A typical Feynman diagram contributing to the emission of a 
photon of momentum $k$ and $n_1$ pions from the initial heavy meson $H_Q$.

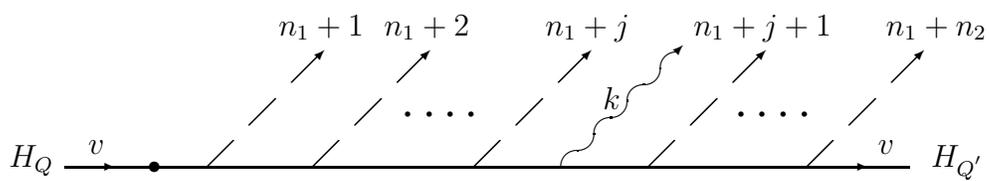
\item[Fig.~10] A typical Feynman diagram contributing to the emission of a 
photon of momentum $k$ and $n_2$ pions from the final heavy meson $H_{Q'}$.

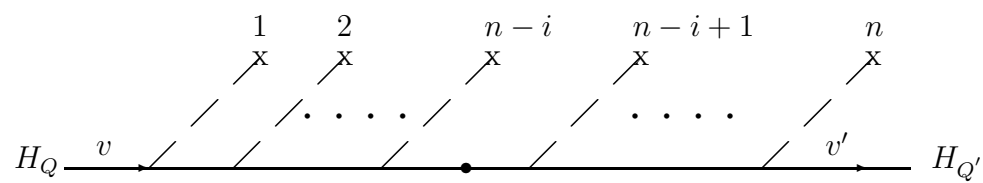
\item[Fig.~11] A typical Feynman diagram contributing to the Isgur-Wise 
function in the presence of an external source represented by the crosses.

\end{description}

\newpage

\textheight 800pt \textwidth 450pt
\pagestyle{empty}
\begin {picture}(0,18000)

\drawline\fermion[\E\REG](15000,15000)[18000]
\put (27000,15000){\circle*{400}}
\drawarrow[\E\ATBASE](\pmidx,\pmidy)
\put(18000,15000){\vector (1,0){0.6}}
\put(30000,15000){\vector(1,0){0.6}}
\put(15010,16000) {(a)}
\put(16900,13500){$\bar{B}$}
\put(23300,13500){$\bar{B}$*}
\put(29200,13500){$D$}
\drawline\scalar[\NE\REG](21000,15000)[3]
\drawarrow[\NE\ATBASE](\pbackx,\pbacky)
\put(25500,20000){$\pi $}
\put(21900,17000){$q$}

\drawline\fermion[\E\REG](15000,3000)[18000]
\put (21000,3000){\circle*{400}}
\drawarrow[\E\ATBASE](\pmidx,\pmidy)
\put(18000,3000){\vector (1,0){0.6}}
\put(30000,3000){\vector(1,0){0.6}}
\put(15010,4000) {(b)}
\put(16900,1500){$\bar{B}$}
\put(23300,1500){$D$*}
\put(29200,1500){$D$}
\drawline\scalar[\NE\REG](27000,3000)[3]
\drawarrow[\NE\ATBASE](\pbackx,\pbacky)
\put(31500,8000){$\pi $}
\put(27900,5000){$q$}

\put(23000,-2500){FIG. 1. }

\drawline\fermion[\E\REG](15000,-11500)[18000]
\put (27000,-11500){\circle*{400}}
\drawarrow[\E\ATBASE](\pmidx,\pmidy)
\put(18000,-11500){\vector (1,0){0.6}}
\put(30000,-11500){\vector(1,0){0.6}}
\put(15010,-10500) {(a)}
\put(16900,-13000){$\bar{B}$}
\put(23300,-13000){$\bar{B}$*}
\put(29200,-13000){$D$*}
\drawline\scalar[\NE\REG](21000,-11500)[3]
\drawarrow[\NE\ATBASE](\pbackx,\pbacky)
\put(25500,-6500){$\pi $}
\put(21900,-9500){$q$}

\drawline\fermion[\E\REG](15000,-23500)[18000]
\put (21000,-23500){\circle*{400}}
\drawarrow[\E\ATBASE](\pmidx,\pmidy)
\put(18000,-23500){\vector (1,0){0.6}}
\put(30000,-23500){\vector(1,0){0.6}}
\put(15010,-22500) {(b)}
\put(16900,-25000){$\bar{B}$}
\put(23300,-25000){$D$*}
\put(29200,-25000){$D$*}
\drawline\scalar[\NE\REG](27000,-23500)[3]
\drawarrow[\NE\ATBASE](\pbackx,\pbacky)
\put(31500,-18500){$\pi $}
\put(27900,-21500){$q$}

\drawline\fermion[\E\REG](15000,-35500)[18000]
\put (21000,-35500){\circle*{400}}
\drawarrow[\E\ATBASE](\pmidx,\pmidy)
\put(18000,-35500){\vector (1,0){0.6}}
\put(30000,-35500){\vector(1,0){0.6}}
\put(15010,-34500) {(c)}
\put(16900,-37000){$\bar{B}$}
\put(23300,-37000){$D$}
\put(29200,-37000){$D$*}
\drawline\scalar[\NE\REG](27000,-35500)[3]
\drawarrow[\NE\ATBASE](\pbackx,\pbacky)
\put(31500,-30500){$\pi $}
\put(27900,-33500){$q$}

\put(23000,-41000){FIG. 2.}
\end{picture}

\newpage

\topmargin 1in
\oddsidemargin 0cm

\begin{picture}(0,18000)

\drawline\fermion[\E\REG](0,15000)[18000]
\put (12000,15000){\circle*{400}}
\drawarrow[\E\ATBASE](\pmidx,\pmidy)
\put(3000,15000){\vector (1,0){0.6}}
\put(15000,15000){\vector(1,0){0.6}}
\put(10,16000) {(a)}
\put(1900,13500){ $\Sigma _b$}
\put(8300,13500){$\Sigma _b$}
\put(14200,13500){$\Sigma _c$}
\drawline\scalar[\NE\REG](6000,15000)[3]
\drawarrow[\NE\ATBASE](\pbackx,\pbacky)
\put(10700,20000){$\pi $}
\put(6800,17000){$q$}

\drawline\fermion[\E\REG](0,3000)[18000]
\put (12000,3000){\circle*{400}}
\drawarrow[\E\ATBASE](\pmidx,\pmidy)
\put(3000,3000){\vector (1,0){0.6}}
\put(15000,3000){\vector(1,0){0.6}}
\put(10,4000) {(b)}
\put(1900,1500){ $\Sigma _b$ }
\put(8300,1500){ $\Sigma _b^*$ }
\put(14200,1500){ $\Sigma _c$  }
\drawline\scalar[\NE\REG](6000,3000)[3]
\drawarrow[\NE\ATBASE](\pbackx,\pbacky)
\put(10700,8000){$\pi $}
\put(6800,5000){$q$}

\drawline\fermion[\E\REG](0,-9000)[18000]
\put (6000,-9000){\circle*{400}}
\drawarrow[\E\ATBASE](\pmidx,\pmidy)
\put(3000,-9000){\vector (1,0){0.6}}
\put(15000,-9000){\vector(1,0){0.6}}
\put(10,-8000) {(c)}
\put(1900,-10500){ $\Sigma _b$}
\put(8300,-10500){$\Sigma _c$}
\put(14200,-10500){$\Sigma _c$}
\drawline\scalar[\NE\REG](12000,-9000)[3]
\drawarrow[\NE\ATBASE](\pbackx,\pbacky)
\put(16700,-4000){$\pi $}
\put(12800,-7000){$q$}

\drawline\fermion[\E\REG](0,-21000)[18000]
\put (6000,-21000){\circle*{400}}
\drawarrow[\E\ATBASE](\pmidx,\pmidy)
\put(3000,-21000){\vector (1,0){0.6}}
\put(15000,-21000){\vector(1,0){0.6}}
\put(10,-20000) {(d)}
\put(1900,-22500){ $\Sigma _b$}
\put(8300,-22500){$\Sigma _c^*$}
\put(14200,-22500){$\Sigma _c$}
\drawline\scalar[\NE\REG](12000,-21000)[3]
\drawarrow[\NE\ATBASE](\pbackx,\pbacky)
\put(16700,-16000){$\pi $}
\put(12800,-19000){$q$}

\put(8000,-26500){FIG. 3.}

\drawline\fermion[\E\REG](30000,15000)[18000]
\put (42000,15000){\circle*{400}}
\drawarrow[\E\ATBASE](\pmidx,\pmidy)
\put(33000,15000){\vector (1,0){0.6}}
\put(45000,15000){\vector(1,0){0.6}}
\put(30010,16000) {(a)}
\put(31900,13500){ $\Sigma _b$}
\put(38300,13500){$\Sigma _b$}
\put(44200,13500){$\Sigma _c^*$}
\drawline\scalar[\NE\REG](36000,15000)[3]
\drawarrow[\NE\ATBASE](\pbackx,\pbacky)
\put(40700,20000){$\pi $}
\put(36800,17000){$q$}

\drawline\fermion[\E\REG](30000,3000)[18000]
\put (42000,3000){\circle*{400}}
\drawarrow[\E\ATBASE](\pmidx,\pmidy)
\put(33000,3000){\vector (1,0){0.6}}
\put(45000,3000){\vector(1,0){0.6}}
\put(30010,4000) {(b)}
\put(31900,1500){ $\Sigma _b$}
\put(38300,1500){$\Sigma _b^*$}
\put(44200,1500){$\Sigma _c^*$}
\drawline\scalar[\NE\REG](36000,3000)[3]
\drawarrow[\NE\ATBASE](\pbackx,\pbacky)
\put(40700,8000){$\pi $}
\put(36800,5000){$q$}

\drawline\fermion[\E\REG](30000,-9000)[18000]
\put (36000,-9000){\circle*{400}}
\drawarrow[\E\ATBASE](\pmidx,\pmidy)
\put(33000,-9000){\vector (1,0){0.6}}
\put(45000,-9000){\vector(1,0){0.6}}
\put(30010,-8000) {(c)}
\put(31900,-10500){ $\Sigma _b$}
\put(38300,-10500){$\Sigma _c$}
\put(44200,-10500){$\Sigma _c^*$}
\drawline\scalar[\NE\REG](42000,-9000)[3]
\drawarrow[\NE\ATBASE](\pbackx,\pbacky)
\put(46700,-4000){$\pi $}
\put(42800,-7000){$q$}

\drawline\fermion[\E\REG](30000,-21000)[18000]
\put (36000,-21000){\circle*{400}}
\drawarrow[\E\ATBASE](\pmidx,\pmidy)
\put(33000,-21000){\vector (1,0){0.6}}
\put(45000,-21000){\vector(1,0){0.6}}
\put(30010,-20000) {(d)}
\put(31900,-22500){ $\Sigma _b$}
\put(38300,-22500){$\Sigma _c^*$}
\put(44200,-22500){$\Sigma _c^*$}
\drawline\scalar[\NE\REG](42000,-21000)[3]
\drawarrow[\NE\ATBASE](\pbackx,\pbacky)
\put(46700,-16000){$\pi $}
\put(42800,-19000){$q$}

\put(38000,-26500){FIG. 4.}

\end{picture}

\newpage
\topmargin -2in

\begin{picture}(0,18000)

\drawline\fermion[\E\REG](15000,-11500)[18000]
\put (27000,-11500){\circle*{400}}
\drawarrow[\E\ATBASE](\pmidx,\pmidy)
\put(18000,-11500){\vector (1,0){0.6}}
\put(30000,-11500){\vector(1,0){0.6}}
\put(15010,-10500) {(a)}
\put(16900,-13000){ $\Sigma _b$}
\put(23300,-13000){$\Lambda _b$}
\put(29200,-13000){$\Lambda _c$}
\drawline\scalar[\NE\REG](21000,-11500)[3]
\drawarrow[\NE\ATBASE](\pbackx,\pbacky)
\put(25500,-6500){$\pi $}
\put(21900,-9500){$q$}

\drawline\fermion[\E\REG](15000,-23500)[18000]
\put (21000,-23500){\circle*{400}}
\drawarrow[\E\ATBASE](\pmidx,\pmidy)
\put(18000,-23500){\vector (1,0){0.6}}
\put(30000,-23500){\vector(1,0){0.6}}
\put(15010,-22500) {(b)}
\put(16900,-25000){ $\Sigma _b $}
\put(23300,-25000){$\Sigma _c$}
\put(29200,-25000){$\Lambda _c$}
\drawline\scalar[\NE\REG](27000,-23500)[3]
\drawarrow[\NE\ATBASE](\pbackx,\pbacky)
\put(31500,-18500){$\pi $}
\put(27900,-21500){$q$}

\drawline\fermion[\E\REG](15000,-35500)[18000]
\put (21000,-35500){\circle*{400}}
\drawarrow[\E\ATBASE](\pmidx,\pmidy)
\put(18000,-35500){\vector (1,0){0.6}}
\put(30000,-35500){\vector(1,0){0.6}}
\put(15010,-34500) {(c)}
\put(16900,-37000){ $\Sigma _b $ }
\put(23300,-37000){$\Sigma _c^*$}
\put(29200,-37000){$\Lambda _c$}
\drawline\scalar[\NE\REG](27000,-35500)[3]
\drawarrow[\NE\ATBASE](\pbackx,\pbacky)
\put(31500,-30500){$\pi $}
\put(27900,-33500){$q$}

\put(23000,-41000){FIG. 5.}

\end{picture}

\newpage

\topmargin 1.5in
\hspace{-2cm}
\setlength{\oddsidemargin}{2cm}
\begin {picture}(0,15000)

\put(20800,3000){FIG. 6}

\put(5000,10000){$H_Q$}
\put(8200,10500){$v$}

\drawline\fermion[\E\REG](7500,10000)[31000]
\put(9600,10000){\vector (1,0){0.6}}
\drawline\scalar[\NE\REG](9600,10000)[3]
\drawarrow[\NE\ATBASE](\pbackx,\pbacky)
\put(\pbackx,15000){$1$}
\drawline\scalar[\NE\REG](12800,10000)[3]
\drawarrow[\NE\ATBASE](\pbackx,\pbacky)
\put(\pbackx,15000){$2$}
\put (15600,12000){\circle*{250}}
\put (16800,12000){\circle*{250}}
\put (18000,12000){\circle*{250}}
\put (19200,12000){\circle*{250}}

\drawline\scalar[\NE\REG](18400,10000)[3]
\drawarrow[\NE\ATBASE](\pbackx,\pbacky)
\put(\pbackx,15000){$i$}
\put (21600,10000){\circle*{400}}
\drawline\scalar[\NE\REG](24000,10000)[3]
\drawarrow[\NE\ATBASE](\pbackx,\pbacky)
\global\advance\pbackx by -800
\put(\pbackx,15000){$i+1$}

\put (26800,12000){\circle*{250}}
\put (28000,12000){\circle*{250}}
\put (29200,12000){\circle*{250}}
\put (30400,12000){\circle*{250}}
\drawline\scalar[\NE\REG](29600,10000)[3]
\drawarrow[\NE\ATBASE](\pbackx,\pbacky)
\global\advance\pbackx by -1100
\put(\pbackx,15000){$n-1$}
\drawline\scalar[\NE\REG](32800,10000)[3]
\drawarrow[\NE\ATBASE](\pbackx,\pbacky)
\put(\pbackx,15000){$n$}
\put(35200,10500){$v$}
\put(36800,10000){\vector (1,0){0.6}}
\put(39200,10000){$H_{Q^{'}}$}

\end{picture}

\noindent
\begin {picture}(0,15000)


\hspace{-1.5cm}

\put (22000,-2000){FIG. 7.}

\put (22800,9000){\oval(32000,6000)}
\put (4500,9000){$H_Q$}
\drawline\scalar[\NE\REG](6300,8500)[1]  %
\drawline\scalar[\SE\REG](6300,9400)[1]  
\put (39600,9000){$H_{Q^{'}}$}
\drawline\scalar[\NE\REG](38300,8500)[1]  %
\drawline\scalar[\SE\REG](38300,9400)[1]  
\put (14000,12400){$Q$}
\put (16400,12000){\vector (1,0){0.6}}
\put (22800,12000){\circle*{400}}
\put (30400,12400){$Q^{'}$}
\put (29200,12000){\vector (1,0){0.6}}

\drawline\scalar[\SE\REG](10800,6000)[3]
\drawarrow[\SE\ATBASE](\pbackx,\pbacky)
\put(\pbackx,100){$1$}
\drawline\scalar[\SE\REG](14000,6000)[3]
\drawarrow[\SE\ATBASE](\pbackx,\pbacky)
\put(\pbackx,100){$2$}

\put (21000,6500){$\overline q$}
\put (22800,6000){\vector (-1,0){0.6}}
\put (21200,2500){\circle*{250}}
\put (22800,2500){\circle*{250}}
\put (24400,2500){\circle*{250}}
\put (26000,2500){\circle*{250}}
\drawline\scalar[\SE\REG](28000,6000)[3]
\drawarrow[\SE\ATBASE](\pbackx,\pbacky)
\global\advance\pbackx by 400
\put(31000,100){$n-1$}
\drawline\scalar[\SE\REG](31200,6000)[3]
\drawarrow[\SE\ATBASE](\pbackx,\pbacky)
\put(\pbackx,100){$n$}

\end{picture}



\vspace{4cm}
\hspace{-2cm}
\noindent
\begin {picture}(0,15000)

\put(22000,5000){FIG. 8.}

\put(5700,10000){$H_Q$}
\put(8000,10500){$v$}

\drawline\fermion[\E\REG](7600,10000)[36000]
\put(9500,10000){\vector (1,0){0.6}}
\drawline\scalar[\NE\REG](9500,10000)[3]
\drawarrow[\NE\ATBASE](\pbackx,\pbacky)
\put(\pbackx,15000){$1$}
\global\advance\pbackx by -2500
\put(\pbackx,9000){$1$}
\drawline\scalar[\NE\REG](12700,10000)[3]
\drawarrow[\NE\ATBASE](\pbackx,\pbacky)
\put(\pbackx,15000){$2$}
\global\advance\pbackx by -3500
\put(\pbackx,9000){$2$}
\global\advance\pbackx by 2300
\put (\pbackx,12000){\circle*{250}}
\global\advance\pbackx by 1200
\put (\pbackx,12000){\circle*{250}}
\global\advance\pbackx by 1200
\put (\pbackx,12000){\circle*{250}}
\global\advance\pbackx by 1200
\put (\pbackx,12000){\circle*{250}}



\drawline\scalar[\NE\REG](18300,10000)[3]
\drawarrow[\NE\ATBASE](\pbackx,\pbacky)
\global\advance\pbackx by -800
\put(\pbackx,15000){$n-i$}
\global\advance\pbackx by -2500
\put(\pbackx,9000){$n-i$}
\put (22800,10000){\circle*{400}}
\put(23500,9000){$n-i$}
%
\drawline\scalar[\NE\REG](26500,10000)[3]
\drawarrow[\NE\ATBASE](\pbackx,\pbacky)
\global\advance\pbackx by -2000
\put(\pbackx,15000){$n-i+1$}
\put(27000,9000){$n-i+1$}
\global\advance\pbackx by 4000
\put (\pbackx,12000){\circle*{250}}
\global\advance\pbackx by 1200
\put (\pbackx,12000){\circle*{250}}
\global\advance\pbackx by 1200
\put (\pbackx,12000){\circle*{250}}
\global\advance\pbackx by 1200
\put (\pbackx,12000){\circle*{250}}


\drawline\scalar[\NE\REG](35300,10000)[3]
\drawarrow[\NE\ATBASE](\pbackx,\pbacky)
\global\advance\pbackx by -1000
\put(\pbackx,15000){$n-1$}
\global\advance\pbackx by -3000
\put(\pbackx,9000){$n-1$}

%
\drawline\scalar[\NE\REG](38100,10000)[3]
\drawarrow[\NE\ATBASE](\pbackx,\pbacky)
\put(\pbackx,15000){$n$}
\global\advance\pbackx by -3000
\put(\pbackx,9000){$n$}
\put(42000,10500){$v$}
\put(42500,10000){\vector (1,0){0.6}}
\put(44400,10000){$H_{Q^{'}}$}

\end{picture}

\newpage

\hspace{-2cm}
\noindent
\begin {picture}(0,15000)

\put(22000,3000){FIG. 9}

\put(5500,10000){$H_Q$}
\put(8800,10500){$v$}

\drawline\fermion[\E\REG](7600,10000)[32000]
\put(10800,10000){\vector (1,0){0.6}}
\drawline\scalar[\NE\REG](10800,10000)[3]
\drawarrow[\NE\ATBASE](\pbackx,\pbacky)
\put(\pbackx,15000){$1$}
\drawline\scalar[\NE\REG](14000,10000)[3]
\drawarrow[\NE\ATBASE](\pbackx,\pbacky)
\put(\pbackx,15000){$2$}
\put (16800,12000){\circle*{250}}
\put (18000,12000){\circle*{250}}
\put (19200,12000){\circle*{250}}
\put (20400,12000){\circle*{250}}

\drawline\scalar[\NE\REG](19600,10000)[3]
\drawarrow[\NE\ATBASE](\pbackx,\pbacky)
\put(\pbackx,15000){$i$}
%

\drawline\photon[\NE\REG](22400,10000)[7]
\drawarrow[\NE\ATBASE](\pbackx,\pbacky)
\global\advance\pmidx by -600
\put(\pmidx,\pmidy){$k$}

\drawline\scalar[\NE\REG](25200,10000)[3]
\drawarrow[\NE\ATBASE](\pbackx,\pbacky)
\put(\pbackx,15000){$i+1$}

\put (28000,12000){\circle*{250}}
\put (29200,12000){\circle*{250}}
\put (30400,12000){\circle*{250}}
\put (31600,12000){\circle*{250}}
\drawline\scalar[\NE\REG](30800,10000)[3]
\drawarrow[\NE\ATBASE](\pbackx,\pbacky)
\put(\pbackx,15000){$n_1$}
\put(34000,10000){\circle*{400}}
\put(36000,10500){$v$}

\put(39000,10000){\vector (1,0){0.6}}

\put(40400,10000){$H_{Q^{'}}$}

\end{picture}

\vspace{2cm}
\hspace{-2cm}
\noindent
\begin {picture}(0,15000)

\put(22000,5000){FIG. 10.}

\put(5500,10000){$H_Q$}

\drawline\fermion[\E\REG](7600,10000)[32000]
\put(8500,10500){$v$}
\put(9500,10000){\vector (1,0){0.6}}
\put(11000,10000){\circle*{400}}
%

\drawline\scalar[\NE\REG](13000,10000)[3]
\drawarrow[\NE\ATBASE](\pbackx,\pbacky)
\global\advance\pbackx by -1500
\put(\pbackx,15000){$n_1+1$}

\drawline\scalar[\NE\REG](17000,10000)[3]
\drawarrow[\NE\ATBASE](\pbackx,\pbacky)
\global\advance\pbackx by -1500
\put(\pbackx,15000){$n_1+2$}

\put (20600,12000){\circle*{250}}
\put (21400,12000){\circle*{250}}
\put (22200,12000){\circle*{250}}
\put (23000,12000){\circle*{250}}
\drawline\scalar[\NE\REG](23100,10000)[3]
\drawarrow[\NE\ATBASE](\pbackx,\pbacky)
\global\advance\pbackx by -1500
\put(\pbackx,15000){$n_1+j$}

\drawline\photon[\NE\REG](26400,10000)[7]
\drawarrow[\NE\ATBASE](\pbackx,\pbacky)
\global\advance\pmidx by -600
\put(\pmidx,\pmidy){$k$}

\drawline\scalar[\NE\REG](29700,10000)[3]
\drawarrow[\NE\ATBASE](\pbackx,\pbacky)
\global\advance\pbackx by -2500
\put(\pbackx,15000){$n_1+j+1$}
\put(38400,10500){$v$}
%
%
%

\global\advance\pbackx by 1800
\put (\pbackx,12000){\circle*{250}}
\global\advance\pbackx by 800
\put (\pbackx,12000){\circle*{250}}
\global\advance\pbackx by 800
\put (\pbackx,12000){\circle*{250}}
\global\advance\pbackx by 800
\put (\pbackx,12000){\circle*{250}}
%

\drawline\scalar[\NE\REG](35700,10000)[3]
\drawarrow[\NE\ATBASE](\pbackx,\pbacky)
\global\advance\pbackx by -1200
\put(\pbackx,15000){$n_1+n_2$}
\put(364000,10500){$\nu$}
\put(38000,10000){\vector (1,0){0.6}}
\put(40400,10000){$H_{Q^{'}}$}

\end{picture}

\vspace{2cm}
\hspace{-2cm}
\noindent
\begin {picture}(0,15000)

\put(22000,3000){FIG. 11}

\put(5700,10000){$H_Q$}
\put(8800,10500){$v$}

\drawline\fermion[\E\REG](7600,10000)[32000]
\put(10800,10000){\vector (1,0){0.6}}
\drawline\scalar[\NE\REG](10800,10000)[3]
\global\advance\pbackx by -300
\global\advance\pbacky by -300
\put(\pbackx,\pbacky){x}
\put(\pbackx,15000){$1$}
\drawline\scalar[\NE\REG](14000,10000)[3]
\global\advance\pbackx by -300
\global\advance\pbacky by -300
\put(\pbackx,\pbacky){x}
\put(\pbackx,15000){$2$}
\put (16800,12000){\circle*{250}}
\put (18000,12000){\circle*{250}}
\put (19200,12000){\circle*{250}}
\put (20400,12000){\circle*{250}}

\drawline\scalar[\NE\REG](19600,10000)[3]
\global\advance\pbackx by -300
\global\advance\pbacky by -300
\put(\pbackx,\pbacky){x}
\put(\pbackx,15000){$n-i$}
\put (22800,10000){\circle*{400}}
\drawline\scalar[\NE\REG](25200,10000)[3]
\global\advance\pbackx by -300
\global\advance\pbacky by -300
\put(\pbackx,\pbacky){x}
\put(\pbackx,15000){$n-i+1$}
\put (29200,12000){\circle*{250}}
\put (30400,12000){\circle*{250}}
\put (31600,12000){\circle*{250}}
\put (32800,12000){\circle*{250}}
\drawline\scalar[\NE\REG](34000,10000)[3]
\global\advance\pbackx by -300
\global\advance\pbacky by -300

\put(\pbackx,\pbacky){x}  
\put(\pbackx,15000){$n$}
\put(36400,10500){$v'$}
\put(38000,10000){\vector (1,0){0.6}}
\put(40400,10000){$H_{Q^{'}}$}

\end{picture}


\end{document}